\documentclass[11pt,runningheads,a4paper]{article}
\usepackage{amsmath}
\usepackage{graphicx}
\usepackage{amssymb}
\usepackage[arrow,matrix]{xy}

\begin{document}
\date{}
\title{Exact solutions for a rotational flow of generalized second grade fluids through a circular cylinder}
\maketitle
\author\begin{center}{Amir Mahmood}\footnote{\copyright Amir Mahmood,\,\,Saifullah,\,\,Qammar Rubab },\,\, Saifullah,\,\, Qammar Rubab
\end{center}
\textbf{Abstract.}{\footnotesize{
\noindent In this note the velocity field and the associated
tangential stress corresponding to the rotational flows of a
generalized second grade fluid within an infinite circular cylinder
are determined by means of the Laplace and Hankel transforms. At
time $t=0$ the fluid is at rest and the motion is produced by the
rotation of the cylinder, around its axis, with the angular velocity
$\Omega t$. The velocity field and the adequate shear stress are
presented under integral and series forms in terms of the
generalized $G$-functions. Furthermore, they are presented as a sum
between the Newtonian solutions and the adequate non-Newtonian
contributions. The corresponding solutions for the ordinary second grade fluid
and Newtonian fluid are obtained as particular cases of our solutions
for $\beta = 1$\,, respectively $\alpha = 0$\, and\, $\beta = 1$\,.}}\\
\section{Introduction}
\indent \indent The motion of a fluid in a rotating or sliding
cylinder is of interest to both theoretical and practical domains.
It is of very important significance to study the mechanism of
viscoelastic fluids flow in many industry fields, such as oil
 exploitation, chemical and food industry and bio-engineering [1].
 Fetecau \emph{et al.} [2] have considered the general case of helical flow
 of an Oldroyd-B fluid and have determined the velocity fields and
 the associated tangential stresses in forms of series in terms of
 Bessel functions. Recently fractional calculus has encountered much
 success in the description of complex dynamics, such as relaxation,
 oscillation, wave and viscoelastic behaviour. Bagley [3], He [4],
 Tan [5] used fractional calculus to handle various
 problems regarding to flow of the second grade fluid. In this note
 we will study the rotational flow of a generalized second grade
 fluid within an infinite circular cylinder of radius $R$. The motion
 is due to the cylinder that at time $t = 0^{+}$, begins to rotate
 around its axis with the angular velocity $\Omega t$.

Exact analytic solutions of this problem are obtained by using Hankel
and Laplace transforms and generalized $G$-functions. Some classical
results can be obtained as special cases of our solutions.
\section{Governing equations}
\indent \indent  The constitutive equation of an incompressible
generalized second grade fluid is given by [4 -- 6]
\begin{eqnarray}
\mathbf{T} = -p\mathbf{I} + \mu \mathbf{A_1} + \alpha_1
\mathbf{A_2} + \alpha_2 \mathbf{A^2_1}\,,                                        
\end{eqnarray}
where $\textbf{T}$ is the Cauchy stress tensor, $-p\textbf{I}$
denotes the indeterminate spherical stress, $\mathbf{\mu}$ is the
coefficient of viscosity, $\alpha_1$ and $\alpha_2$ are the normal
stress moduli and $\mathbf{A_1}$ and $\mathbf{A_2}$ are the kinematic
tensors defined through
\begin{eqnarray}
\mathbf{A_1}= grad\, \mathbf{v} + (grad\, \mathbf{v})^T\,,              
\end{eqnarray}
\begin{eqnarray}
\mathbf{A_2}= D^{\beta}_t \mathbf{A_1} + \mathbf{A_1}(grad\,              
\mathbf{v}) + (grad\, \mathbf{v})^T \mathbf{A_1}\,.
\end{eqnarray}

In the above relations $\textbf{v}$ is the velocity , the superscript
$T$ denotes the transpose operator, and $D^{\beta}_t$ is the
 Riemann - Liouville fractional derivative operator defined by [7]
\begin{eqnarray}
D^{\beta}_t f(t) =
\frac{1}{\Gamma(1-\beta)}\frac{d}{dt}\int_{0}^{t}\frac{f(\tau)}{(t-\tau)^{\beta}}d\tau         
; \,\,\,\,\,\,\,0 < \beta\leq 1\,,
\end{eqnarray}
where $\Gamma(\cdot)$ is the Gamma function. For $\beta = 1$
the generalized model reduces to classical model of second
grade fluid because $D^1_t f = df/dt$.\\ Since the
fluid is incompressible, it can undergo only isochoric motions and hence
\begin{eqnarray}
div\,\textbf{v} = tr\,\mathbf{A_1}= 0.                                               
\end{eqnarray}
If this model is required to be compatible with thermodynamics, then
the material moduli must meet the following restrictions [8]
\begin{eqnarray}
\mu \geq 0\,\,, \,\,\,\,\, \alpha_1 \geq 0\,\,\,\,\,                           
\hbox{and}\,\,\,\,\,\ \alpha_1 + \alpha_2 = 0\,.
\end{eqnarray}

In cylindrical coordinates $(r, \theta, z)$, the rotational flow
velocity is given by [2, 6]
\begin{eqnarray}
\textbf{v} = \textbf{v}(r, t) = \omega(r, t)\mathbf{e_\theta},                                              
\end{eqnarray}
where $\mathbf{e_\theta}$ is the unit vector in the $\theta$ direction. For such
flows the constraint of incompressibility is automatically
satisfied.\\ Introducing (7) into constitutive equation, we find
that
\begin{eqnarray}
\tau(r, t)= (\mu + \alpha_1 D^\beta_t)(\frac{\partial}{\partial r} -                            
\frac{1}{r})\omega(r, t),
\end{eqnarray}
where $\tau(r, t) = S_{r\theta}(r, t)$ is the shear stress which
is different of zero.
The last equation together with the equations of motion lead to the governing equation
\begin{eqnarray}
\frac{\partial\omega(r, t)}{\partial t} = (\nu + \alpha
D^\beta_t)(\frac{\partial^2}{\partial r^2} +
\frac{1}{r}\frac{\partial}{\partial r} - \frac{1}{r^2})\omega(r,                        
t)\,,\,\,\,\,\,\,\, r\in(0, R)\,,\,\,\, t > 0\,,
\end{eqnarray}
where $\nu = \mu/\rho$ is the kinematic
viscosity, $\rho$ is the constant density of the fluid and $\alpha =
\alpha_1/\rho$.
\section{On the rotational flow through an infinite circular cylinder}
\indent \indent Let us consider an incompressible generalized second grade fluid at
rest in an infinite circular cylinder of radius $R$. At time zero, the
cylinder suddenly begins to rotate about its axis with the angular
velocity $\Omega t$. Owing to the shear, the fluid is gradually
moved, and its velocity is (7) and governing equation is (9). The
appropriate initial and boundary conditions are
\begin{eqnarray}
\omega(r, 0)=0\,;\,\,\,\,\,r\in[0, R)\,,\,\,\,\,\,\,\,\,\,\,\,\, \omega(R, t)=R\Omega               
t\,;\,\,\,\,\,\, t \geq 0\,.
\end{eqnarray}
To solve this problem we shall use as in [6, 9] the Laplace and
Hankel transforms.
\subsection{Calculation of the velocity field}
\indent \indent Applying the Laplace transform to Eqs. (9) and (10) and using the
Laplace transform formula for sequential fractional derivatives [7],
we obtain
\begin{eqnarray}
(\nu+\alpha q^{\beta})\bigg(\frac{\partial^2}
{\partial r^2}+\frac{1}{r}\frac{\partial}{\partial r}
-\frac{1}{r^2}\bigg)\overline{\omega}(r,q)- q\overline{\omega}(r,q)= 0\,,                
\end{eqnarray}
where the image function $\overline{\omega}(r,
q)=\int_0^\infty\omega(r,t)e^{-qt}dt$ of $\omega(r, t)$ has to
satisfy the condition
\begin{eqnarray}
\overline{\omega}(R, q) = \frac{R\Omega}{q^2}\,,                                  
\end{eqnarray}
$q$ being the transform parameter.\\
In the following we denote by
\begin{eqnarray}
\overline{\omega}_{{}_H}(r_{1n},q)=\int_0^Rr\overline{\omega}(r,q)J_1(rr_{1n})dr\,,             
\end{eqnarray}
the Hankel transform of $\overline{\omega}(r,q)$ , where
$J_1(\cdot)$ is the Bessel function of first kind of order one and
$r_{1n}\,, n = 1, 2, 3, ...$  are the positive roots of the
transcendental equations $J_1(Rr)=0$.\\ Multiplying now both sides
of Eq. (11) by $rJ_1(rr_{1n})$, integrating with respect to $r$ from
$0$ to $R$ and taking into account the condition (12) and the
equality
\begin{eqnarray}
\nonumber\int_0^Rr\Big[\frac{\partial^2\overline{\omega}(r,q)}{\partial             
r^2}+ \frac{1}{r} \frac{\partial\overline{\omega}(r,q)}{\partial
r}-\frac{\overline{\omega}(r,q)}{ r^2}\Big]J_1(rr_{1n})dr =\\=
Rr_{1n}J_2(Rr_{1n})
\overline{\omega}(R,q)- r^2_{1n}\overline{\omega}_{H}(r_{1n},q)\,,
\end{eqnarray}
we find that
\begin{eqnarray}
\overline{\omega}_{H}(r_{1n},q)= \Omega
R^2r_{1n}J_2(Rr_{1n})\frac{\nu +\alpha q^\beta}{q^2\Big[q + \alpha             
r^2_{1n}q^\beta + \nu r^2_{1n}\Big]}\,.
\end{eqnarray}
Now, for a more suitable presentation of the final results, we
rewrite Eq. (15) in the following equivalent form
\begin{eqnarray}
\overline{\omega}_{H}(r_{1n},q)= \overline{\omega}_{1H}(r_{1n},q)+
\overline{\omega}_{2H}(r_{1n},q)+ \overline{\omega}_{3H}(r_{1n},q)\,,             
\end{eqnarray}
where
\begin{eqnarray}
\overline{\omega}_{1H}(r_{1n},q) = \frac{\Omega R^2}{q^2 r_{1n}}J_2(Rr_{1n})\,,             
\end{eqnarray}
\begin{eqnarray}
\overline{\omega}_{2H}(r_{1n},q) = -\frac{\Omega R^2
J_2(Rr_{1n})}{\nu r^3_{1n}}\bigg(\frac{1}{q} - \frac{1}{q + \nu             
r^2_{1n}}\bigg)
\end{eqnarray}
and
\begin{eqnarray}
\overline{\omega}_{3H}(r_{1n},q) = \alpha \Omega R^2 r_{1n}J_2(R             
r_{1n})\frac{1}{q + \nu r^2_{1n}}\frac{q^{\beta - 1}}{\Big[q +
\alpha r^2_{1n} q^\beta + \nu r^2_{1n}\Big]}\,.
\end{eqnarray}
Using the formula
\begin{eqnarray}
\int_0^Rr^2J_1(rr_{1n})dr = \frac{R^2}{r_{1n}}J_2(Rr_{1n})\,,             
\end{eqnarray}
we get that inverse Hankel transform of the function
$\overline{\omega}_{1H}(r_{1n},q)$ is the function
\begin{eqnarray}
\overline{\omega}_1(r,q) = \frac{\Omega r}{q^2}\,.                   
\end{eqnarray}
The inverse Hankel transforms of the functions
$\overline{\omega}_{kH}(r_{1n},q)\,,\,\,\,\,k = 2, 3$ are the
functions
\begin{eqnarray}
\overline{\omega}_{kH}(r,q) = \frac{2}{R^2}\sum^{\infty}_{n =                     
1}\frac{J_1(r r_{1n})}{J^2_2 (R
r_{1n})}\overline{\omega}_{kH}(r_{1n},q)\,.
\end{eqnarray}
Introducing Eqs. (21) and (22) into Eq. (16) we find that the
Laplace transform $\overline{\omega}(r,q)$ has the form
\begin{eqnarray}
\nonumber\overline{\omega}(r,q) = \frac{\Omega r}{q^2}-\frac{2\Omega                      
}{\nu}\sum^{\infty}_{n = 1}\frac{J_1(r r_{1n})}{r^3_{1n}J_2 (R
r_{1n})}\bigg(\frac{1}{q} - \frac{1}{q + \nu r^2_{1n}}\bigg) +\\
+2\alpha\Omega\sum^{\infty}_{n = 1}\frac{r_{1n}J_1(r r_{1n})}{J_2 (R
r_{1n})}\frac{1}{q + \nu r^2_{1n}}\frac{q^{\beta - 1}}{\Big[q +
\alpha r^2_{1n}q^\beta +\nu r^2_{1n}\Big]}\,.
\end{eqnarray}
To obtain the velocity field $\omega(r, t) =
L^{-1}\{\overline{\omega}(r,q)\}$ we will apply the discrete inverse
Laplace transform method [6, 7, 9]. For this we use the expansion
\begin{eqnarray}
\nonumber F(q) = \frac{q^{\beta - 1}}{q+\alpha r^2_{1n}q^\beta+\nu                       
r_{1n}^2} = \frac{q^{-1}}{(q^{1-\beta}+\alpha r_{1n}^2) + \nu
r^2_{1n}q^{-\beta}} =\\= \sum^{\infty}_{k = 0}(-\nu r^2_{1n})^k
\frac{q^{-\beta k-1}}{\Big(q^{1-\beta}+\alpha
r^2_{1n}\Big)^{k+1}}\,.
\end{eqnarray}
Introducing (24) into (23), applying the discrete inverse Laplace
transform and using the following properties
\begin{eqnarray*}
L^{-1}\{F_1(q)F_2(q)\}=(f_1 *f_2)(t) = \int_0^t f_1(t -
s)f_2(s)ds\,,
\end{eqnarray*}
\begin{eqnarray}
{}                                                                          
\end{eqnarray}
where
\begin{eqnarray*}
f_k(t) = L^{-1}\{F_k(q)\}\,,\,\,\,\,\,\,\,\,k = 1, 2\,,
\end{eqnarray*}
\begin{eqnarray}
L^{-1}\Big\{\frac{q^b}{(q^a -d)^c}\Big\} = G_{a, b, c}(d,                                             
t)\,,\,\,\,\,\, Re\,(ac - b)
> 0\,,
\end{eqnarray}
and [10]
\begin{eqnarray}
G_{a,b,c}(d,t)=\sum_{j=0}^\infty\frac{d^j \Gamma(c + j)}
{\Gamma(c)\Gamma(j + 1)}\frac{t^{(c+j)a-b-1}}{\Gamma[(c+j)a-b]}\,,                                             
\end{eqnarray}
are the generalized G-functions, we find for $\omega(r, t)$ the
expression
\begin{eqnarray*}
\omega(r,t)=\omega_{{}_N}(r,t)+2\alpha\,\Omega\sum_{n=1}^\infty
\frac{r_{1n}J_1(rr_{1n})}{J_2(Rr_{1n})} \sum_{k=0}^\infty\bigg(-\nu
\,r^2_{1n}\bigg)^k\,\times
\end{eqnarray*}
\begin{eqnarray}
\times\int_0^texp[-\nu r_{1n}^2(t-s)]G_{1-\beta,\,-\beta k-1,\,                             
k+1}\bigg(-\alpha r^2_{1n}, s\bigg)ds\,,
\end{eqnarray}
where [2, Eq. (4.5)]
\begin{eqnarray}
\omega_{{}_N}(r,t)=r\Omega t -\frac{2\Omega}{\nu}\sum_{n=1}^\infty                      
\frac{J_1(rr_{1n})}{r_{1n}^3J_2(Rr_{1n})}\big[1-
exp(-\nu r_{1n}^2t)\big]
\end{eqnarray}
is the similar solution for Newtonian fluids, performing same
motion.
\subsection{Calculation of the shear stress}
\indent \indent Applying the Laplace transform to Eqs. (8) we find that
\begin{eqnarray}
\overline{\tau}(r,q)=(\mu + \alpha_1 q^\beta)(\frac{\partial}{\partial r}-               
\frac{1}{r})\overline{\omega}(r,q)\,.
\end{eqnarray}
The image function $\overline{\omega}(r,q)$ can be obtained using
Eqs. (27)-(29) and the formula
\begin{eqnarray}
L\Big\{\frac{t^a}{\Gamma (a + 1)}\Big\} = \frac{1}{q^{a +          
1}}\,,\,\,\,\,\,\,\,\,\, a > -1\,.
\end{eqnarray}
Consequently, applying the Laplace transform to Eq. (28),
differentiating the result with respect to $r$ and using the
identity
\begin{eqnarray}
rJ'_1(rr_{1n})-J_1(rr_{1n})=-rr_{1n}J_2(rr_{1n})\,,    
\end{eqnarray}
we find that
\begin{eqnarray*}
\frac{\partial\overline{\omega}}{\partial
r}-\frac{\overline{\omega}}{r}
=\frac{2\Omega}{\nu}\sum_{n=1}^\infty\frac{J_2(rr_{1n})}{r_{1n}^2J_2(Rr_{1n})}
\bigg(\frac{1}{q}-\frac{1}{q+\nu r_{1n}^2}\bigg) -
\end{eqnarray*}
\begin{eqnarray*}
-2\alpha\Omega \sum_{n=1}^\infty
\frac{r_{1n}^2J_2(rr_{1n})}{J_2(Rr_{1n})} \sum_{k,j=0}^\infty\frac{
\bigg(-\nu \,r^2_{1n}\bigg)^k\bigg(-\alpha \,r^2_{1n}\bigg)^j
\Gamma(k + j + 1)}{\Gamma(k + 1)\Gamma(j + 1)}\times
\end{eqnarray*}
\begin{eqnarray}
\times\frac{1}{q+\nu r_{1n}^2} \frac{1}{q^{k+(1-\beta)(j+1)+1}}\,.                     
\end{eqnarray}
Introducing (33) into (30) we get
\begin{eqnarray*}
\overline{\tau}(r,q)=2\rho\Omega\sum_{n=1}^\infty
\frac{J_2(rr_{1n})}{r^2_{1n}J_2(Rr_{1n})}\bigg(\frac{1}{q} -
\frac{1}{q + \nu r^2_{1n}}\bigg)+ 2\alpha_1\Omega
\sum^\infty_{n=1}\frac{J_2(rr_{1n})}{J_2(Rr_{1n})}\frac{q^{\beta
-1}}{q + \nu r^2_{1n}}-
\end{eqnarray*}
\begin{eqnarray*}
-2\alpha\Omega
\sum^\infty_{n=1}\frac{r_{1n}^2J_2(rr_{1n})}{J_2(Rr_{1n})}
\sum_{k,j=0}^\infty\frac{\bigg(-\nu \,r^2_{1n}\bigg)^k\bigg(-\alpha
r^2_{1n}\bigg)^j\Gamma(k+j+1)}{\Gamma(k+1)\Gamma(j+1)}\times
\end{eqnarray*}
\begin{eqnarray}
\times\bigg\{\frac{1}{q+\nu r^2_{1n}}\bigg[
\frac{\mu}{q^{k+(1-\beta)(j+1)+1}}-\frac{\nu\alpha_1
r^2_{1n}}{q^{k+3+(1-\beta)j
-2\beta}}\bigg]+\frac{\alpha_1}{q^{k+3+(1-\beta)j-2\beta}}\bigg\}\,.                  
\end{eqnarray}
Applying the inverse Laplace transform to Eq. (34), we find that
the shear stress $\tau(r,t)$ has the form
\begin{eqnarray*}
\tau(r,t)=\tau_{{}_{N}}(r,t)+2\alpha_1\Omega\sum_{n=1}^\infty
\frac{J_2(rr_{1n})}{J_2(Rr_{1n})}G_{1, \beta -1, 1}(- \nu r^2_{1n},
t)-
\end{eqnarray*}
\begin{eqnarray*}
-2\alpha\Omega\sum_{n=1}^\infty
\frac{r^2_{1n}J_2(rr_{1n})}{J_2(Rr_{1n})}\sum_{k,j=0}^\infty\frac{
\bigg(-\nu \,r^2_{1n}\bigg)^k\bigg(-\alpha \,r^2_{1n}\bigg)^j
\Gamma(k + j + 1)}{\Gamma(k + 1)\Gamma(j + 1)}\times
\end{eqnarray*}
\begin{eqnarray*}
\times\int^t_0exp[-\nu r_{1n}^2(t-s)]\bigg\{\frac{\mu
s^{k+(1-\beta)(j+1)}}{\Gamma[k+(1-\beta)(j+1)+1]} -
\end{eqnarray*}
\begin{eqnarray*}
-\frac{\nu\alpha_1
r^2_{1n}s^{k+2+(1-\beta)j-2\beta}}{\Gamma[k+3+(1-\beta)j-2\beta]}\bigg\}ds\,-
2\alpha\Omega\sum_{n=1}^\infty
\frac{r^2_{1n}J_2(rr_{1n})}{J_2(Rr_{1n})}\times
\end{eqnarray*}
\begin{eqnarray}
\times\sum_{k,j=0}^\infty\frac{
\bigg(-\nu \,r^2_{1n}\bigg)^k\bigg(-\alpha \,r^2_{1n}\bigg)^j
\Gamma(k + j + 1)}{\Gamma(k + 1)\Gamma(j + 1)}\frac{\alpha_1
t^{k+2+(1-\beta)j-2\beta}}{\Gamma[k+3+(1-\beta)j-2\beta]}                                    
\end{eqnarray}
where [2, Eq. (5.3) for $\alpha=0$]
\begin{eqnarray}
\tau_{{}_{N}}(r,t)=2\rho\Omega\sum_{n=1}^\infty
\frac{J_2(rr_{1n})}{r_{1n}^2J_2(Rr_{1n})}\big[1-
exp(-\nu r_{1n}^2t)\big]\,,                                                          
\end{eqnarray}
is the shear stress corresponding to a Newtonian fluid performing
the same motion.
\section{Limiting cases}
\indent \indent Making $\beta = 1$ into Eqs. (28), we obtain the velocity field
\begin{eqnarray*}
\omega(r,t)=\omega_{{}_N}(r,t)+2\alpha\Omega\sum_{n=1}^\infty
\frac{r_{1n}J_1(rr_{1n})}{J_2(Rr_{1n})} \sum_{k=0}^\infty\bigg(-\nu
\,r^2_{1n}\bigg)^k\times
\end{eqnarray*}
\begin{eqnarray}
\times\int_0^texp[-\nu r_{1n}^2(t-s)]G_{0,-k-1,k+1}
\bigg(-\alpha r^2_{1n},s\bigg)ds\,,                                                  
\end{eqnarray}
corresponding to an ordinary second grade fluid, performing the same
motion. Similarly, from (35), we obtain the shear stress
\begin{eqnarray*}
\tau(r,t)=\tau_{{}_{N}}(r,t)+2\alpha_1\Omega\sum_{n=1}^\infty
\frac{J_2(rr_{1n})}{J_2(Rr_{1n})}G_{1,\, 0, 1}(-\nu r^2_{1n},
t)-
\end{eqnarray*}
\begin{eqnarray*}
-2\alpha\Omega\sum_{n=1}^\infty
\frac{r^2_{1n}J_2(rr_{1n})}{J_2(Rr_{1n})}\sum_{k, j=0}^\infty\frac{(-\nu \,r^2_{1n})^k(-\alpha
r^2_{1n})^j\Gamma(k+j+1)} {\Gamma(k+1)\Gamma(j+1)}\times
\end{eqnarray*}
\begin{eqnarray*}
\times\int_0^texp[-\nu
r_{1n}^2(t-s)](\mu + \nu\alpha_1 r^2_{1n})\frac{s^k}{\Gamma(k +
1)}ds-
\end{eqnarray*}
\begin{eqnarray}
-2\alpha\alpha_1\Omega
\sum_{n=1}^\infty\frac{r_{1n}^2J_2(rr_{1n})}{J_2(Rr_{1n})}
\sum_{k,
j=0}^\infty\frac{(-\nu \,r^2_{1n})^k(-\alpha r^2_{1n})^j
\Gamma(k+j+1)}{\Gamma(k+1)\Gamma(j+1)}\frac{t^k}{\Gamma(k+1)}\,,                           
\end{eqnarray}
corresponding to an ordinary second grade fluid, performing the same
motion.\\ The above relations can be simplified if we use the
following relations:
\begin{eqnarray}
\nonumber G_{0, -k-1, k+1}(-\alpha r^2_{1n}, s) =
\frac{s^k}{\Gamma(k+1)}\sum_{j=0}^\infty\frac{(-\alpha r^2_{1n})^j
\Gamma(k + j + 1)}{\Gamma(k+1)\Gamma(j+1)} =\\= \frac{s^k}{\Gamma(k                 
+ 1)}(1 + \alpha r^2_{1n})^{-(k+1)}\,,
\end{eqnarray}
\begin{eqnarray*}
\sum_{k=0}^\infty(-\nu r^2_{1n})^k G_{0, -k-1, k+1}(-\alpha r^2_{1n}, s) =
\frac{1}{1 + \alpha r^2_{1n}}\sum_{k=0}^\infty\frac{1}{k!}
\bigg(-\frac{\nu r^2_{1n} s}{1 + \alpha r^2_{1n}}\bigg)^k =
\end{eqnarray*}
\begin{eqnarray}
= \frac{1}{1 + \alpha r^2_{1n}} exp\bigg({-\frac{\nu r^2_{1n} s}{1 + \alpha r^2_{1n}}}\bigg)\,,        
\end{eqnarray}
and
\begin{eqnarray}
G_{1, 0, 1}(-\nu r^2_{1n}, t) = exp\bigg({-\nu r^2_{1n} t}\bigg)\,.                        
\end{eqnarray}
As a result, we find the velocity field and the adequate shear stress under simplified forms
\begin{eqnarray}
\omega(r,t)=r\Omega t-\frac{2\Omega}{\nu}\sum_{n=1}^\infty\frac{J_1(rr_{1n})}{r^3_{1n}J_2(Rr_{1n})}       
\bigg[1-exp\bigg(-\frac{\nu r_{1n}^2}{1+\alpha r_{1n}^2}t\bigg)\bigg]
\end{eqnarray}
and
\begin{eqnarray}
\tau(r,t)=2\rho\Omega \sum_{n=1}^\infty\frac{J_2(rr_{1n})}{r^2_{1n}J_2(Rr_{1n})}                  
\bigg[1-\frac{1}{1+\alpha r_{1n}^2}exp\bigg(-\frac{\nu r_{1n}^2}{1+\alpha r_{1n}^2}t\bigg)\bigg]
\end{eqnarray}
which are identical to Eqs. (5.1) and (5.3) from [2].

If in Eqs. (42) and (43), we make $\alpha = 0$\,, then the corresponding solutions of the Newtonian fluids are recovered.
\section{Conclusion}
\indent \indent In this note, the velocity field and the adequate shear stress
corresponding to the rotational flow induced by an infinite circular cylinder in an  incompressible
generalized second grade fluid, have been determined using Hankel and Laplace transforms. The motion is produced by the circular cylinder that at the initial moment begins to rotate around its axis with a constant angular acceleration. The solutions that have been obtained, written under integral and series forms in terms of generalized $G$-function, satisfy all imposed initial and boundary conditions. Furthermore, they are presented as a sum between the Newtonian solutions and the adequate non-Newtonian contributions. In the special case when $\beta=1$\,, or $\beta=1$ and $\alpha=0$\,, the corresponding solutions for ordinary second grade fluid and Newtonian fluid, respectively, performing same motion, are obtained.



\vskip2cm
AMIR MAHMOOD\\
Abdus Salam School of Mathematical Sciences,\\
GCU, Lahore\\
E-mail: \emph{amir4smsgc@gmail.com}\\\\
SAIFULLAH\\
Abdus Salam School of Mathematical Sciences,\\
GCU, Lahore\\
E-mail:\emph{ saifullahkhalid75@yahoo.com}\\\\
QAMMAR RUBAB\\
FAST National University of Computer\\ \& Emerging Sciences,
Lahore\\
E-mail: \emph{Rubabqammar@gmail.com}

\end{document}